\makeatletter
\def\input@path{{./library/}}
\makeatother

\documentclass[12pt]{iopart}
\usepackage{graphicx}
\usepackage{xcolor}
\expandafter\let\csname equation*\endcsname\relax
\expandafter\let\csname endequation*\endcsname\relax
\usepackage{amsmath}
\usepackage{makecell}
\usepackage[colorlinks=true,
            linkcolor=blue,
            citecolor=blue,
            urlcolor=blue]{hyperref}

\begin{document}

\title[]{Variational Autoencoders for At-Source Data Reduction and Anomaly Detection in High Energy Particle Detectors}

\author{Alexander Yue$^{1}$, Haoyi Jia$^{1,2}$, Julia Gonski$^{2}$}
\address{$^{1}$ Stanford University,  450 Jane Stanford Way,  Stanford, CA 94305, United States of America}
\address{
    $^{2}$ SLAC National Accelerator Laboratory,
    2575 Sand Hill Road MS 95,
    Menlo Park, CA 94025, United States of America
}

\ead{jgonski@slac.stanford.edu}
\vspace{10pt}




\begin{abstract}

Detectors in next-generation high-energy physics experiments face several daunting requirements, such as high data rates, damaging radiation exposure, and stringent constraints on power, space, and latency. 
To address these challenges, machine learning in readout electronics can be leveraged for smart detector designs, enabling intelligent inference and data reduction at-source. 
Variational autoencoders (VAEs) offer a variety of benefits for front-end readout; an on-sensor encoder can perform efficient lossy data compression while simultaneously providing a latent space representation that can be used for anomaly detection. 
Results are presented from low-latency and resource-efficient VAEs for front-end data processing in a futuristic silicon pixel detector. 
Encoder-based data compression is found to preserve good performance of off-detector analysis while significantly reducing the off-detector data rate as compared to a similarly sized data filtering approach. 
Furthermore, the latent space information is found to be a useful discriminator in the context of real-time sensor defect monitoring. 
Together, these results highlight the multifaceted utility of autoencoder-based front-end readout schemes and motivate their consideration in future detector designs.

\end{abstract}


\section{Introduction}
\label{sec:intro}

Research in high-energy collider physics requires the design and construction of high-performance particle detectors to reconstruct and identify all standard model particles produced in collision events, while also providing sensitivity to the potential presence of new, beyond-the-standard-model physics. 
Equally as important are the systems of readout electronics that convert the signals of particles passing through detector material into data usable for analysis of collision phenomena. 
It is crucial that both detector and readout concepts evolve to keep up with community priorities for future colliders. 
The 2023 Particle Physics Project Prioritization Panel (P5) outlines dual goals of a precision $e^+e^-$ Higgs factory to be constructed in the coming decades, and R\&D for a 10 TeV parton-center-of-mass discovery accelerator in the further future~\cite{p5_2023}. 

Achieving the physics goals of these facilities, including unprecedented precision measurements of the Higgs boson and novel coverage of new physics hypotheses, requires the development and application of cutting-edge technology.
These advances must address the new challenges of future collider environments, such as high data density, low latency, radiation dose, and material budget constraints.
The incorporation of artificial intelligence by way of machine learning (ML) deployed throughout collider trigger and data acquisition (DAQ) schemes can help mitigate some of these challenges, and thus has been highlighted as a key R\&D priority by the community~\cite{doe_brn, apresyan2023detector}.

Among the more challenging areas of applying ML to collider DAQ is at the front-end.
These electronics systems must process detector signals on the timescale of $\mathcal{O}$(ns) bunch crossings while also enduring the high radiation dose of the experimental cavern. 
Implementation of ML at this ``edge", or at the source of data, is a rapidly growing area of interest and requires special considerations of both computing hardware and algorithms~\cite{duarte2022fastmlsciencebenchmarksaccelerating, carini2022smartsensor}.

The choice of hardware for the deployment of edge ML models is crucial for operation in the collider context. 
Hardware must combine the benefits of an application-specific integrated circuit (ASIC), namely low power, fast latencies, and radiation tolerance, with the reconfigurability necessary for a robust ML implementation. 
Previous work in ML on ASICs has implemented spiking neural network models for edge data filtering~\cite{kulkarni2023onsensordatafilteringusing}. 
Another technology that fulfills these requirements is the embedded field programmable gate array (eFPGA), which offers fully reconfigurable digital logic within an ASIC. 
eFPGAs have recently been studied for application to detector readout, such as for silicon tracker data processing~\cite{Gonski_2024} and silicon photomultiplier readout~\cite{Johnson2024_ngc_eFPGAs}.

Parallel to the choice of hardware platform is the study of intelligent algorithms that can be compressed to deliver robust performance amid latency and resource constraints, referred to as ``fast ML"~\cite{10.3389/fdata.2022.787421}.
ML algorithms that can reduce the off-detector data rate would allow for more efficient data transmission while reducing the material budget of data cables in the sensitive detector regions. 
For example, classification and subsequent filtering of high-momentum particles from low-momentum pileup~\cite{yoo2023_on_sensor_filtering} and at-source inference of track parameters~\cite{dickinson2023_on_sensor_inference} have been studied in the context of futuristic ``smart pixel'' simulations~\cite{Dickinson2022_Smart_Pixel_Dataset}. 

The autoencoder is a common ML architecture that can provide a variety of benefits for front-end data analysis and has been studied for high-energy physics readout ASICs~\cite{9447722, Shenoy_2023}.
This work focuses on the variational autoencoder (VAE)~\cite{kingma2022autoencodingvariationalbayes}, which consists of two parts: an encoder that performs a lossy compression of inputs into a latent space of Gaussian distributions defined by their means and variances, and a decoder that samples from this latent space to generate an output of the original dimensionality. 
The model is trained with a loss that includes the reconstruction error of the output to the input, motivating the model to produce increasingly realistic outputs, and a Kullback-Leibler divergence (KLD) term representing the measure of the how the Latent space distribution is different from the Gaussian prior.
To minimize this loss function, the VAE learn to extract the most salient features of the input datasets to be able to generate realistic outputs; this knowledge is embedded within the latent space construction. 

VAEs offer two functionalities useful for on-detector readout: compression, namely by reducing data at-source to the latent space dimensionality for transmission followed by decompression off-detector, and anomaly detection, or the use of latent space information to determine whether an event is compatible with the expected background~\cite{BELIS2024100091}. 
In this work, VAEs are presented that can achieve real-time data compression and anomaly detection with realistic resource usage and latency for front-end implementation in future collider detectors. 
These findings extend prior work by demonstrating that commonly used ML models for data compression can inherently also perform anomaly detection, without requiring additional optimization or processing resources.


\section{Methodology}
\label{sec:methods}

A key component of many high energy physics detectors is the system of silicon pixel sensors that detect the propagation of charged particles from the collision origin. 
The spatial position of deposited charge across layers of silicon pixels can allow for reconstruction of a charged particle's trajectory.
From this trajectory, key information about the particle's true kinematics can be inferred that is ultimately used in a full reconstruction of the collider event. 

\begin{figure}[!htbp]
\centering
\includegraphics[width=0.4\textwidth]{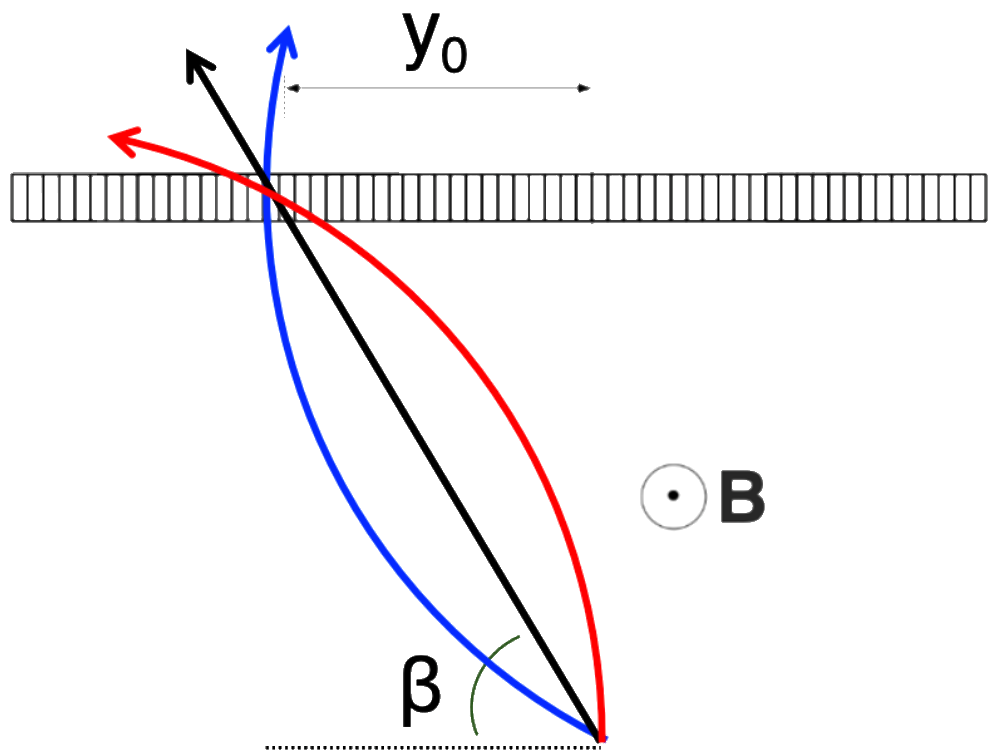}
\caption{
    A visual representation of the simulated detector in the ``smart pixel" dataset \cite{Dickinson2022_Smart_Pixel_Dataset}. 
    In this graphic, a slice along the $y$ axis of the pixel array is shown, where $\beta$ is the track incident angle and the magnetic field $B$ runs along the $x$ direction. 
    The colored red and blue paths represent the curved path of a low momentum pion under the influence of a strong magnetic field, whereas the black path represents the path of a high momentum pion which is less affected by the magnetic field.
    For this detector setup, the relevant factors for determining $p_\text{T}$ are the charge depositions over time of the 13 pixels in the $y$ direction around the particle-sensor intersection, in addition to $y_0$, the offset of the intersection point from the collision origin in the $y$ direction.
}
\label{fig:pixel_sensor}
\end{figure}

The VAE models developed here use a simulated dataset of charged particles passing through a futuristic pixel sensor~\cite{Dickinson2022_Smart_Pixel_Dataset}.
This dataset contains over 500,000 tracks produced by charged pions in a 3.8 T magnetic field. 
The kinematic properties of these tracks are taken from track kinematics fitted from proton collision data events collected by the CMS detector~\cite{CMS:2008xjf} during Run 2 of the Large Hadron Collider (LHC).
The simulated detector is an region of $21 \times 13$ pixels of size $50 \mu\text{m} \times 12.5 \mu\text{m}$ located at a radius of 30 mm from the beamline. For each particle track, the charges deposited within this pixel region at 8 time slices with intervals of 200 ps are recorded. 
As the magnetic field is parallel to the $x$ direction, only information in the $y$ direction is sensitive to the track's transverse momentum $p_\text{T}$.
A visual representation of the pixel array accompanied by a description of the readout format is found in Figure \ref{fig:pixel_sensor}.

To accurately assess the potential resources and latency of running a VAE in real-time during data collection, VAE algorithms are first developed in software and then synthesized to register-transfer level for hardware deployment. 
Model synthesis is performed using the framework \textsc{hls4ml}~\cite{Duarte:2018ite, fastml_hls4ml}, an open-source framework for translating machine learning algorithms into hardware implementations.
Such synthesized models must accommodate the key constraints of on-detector application.
Specifically, the 40 MHz peak bunch crossing rate of the LHC imposes a $25$ ns latency requirement for single bunch crossing inference, and power density limitations on the detector impose constraints on the logical resources required by the algorithm. 
Using different levels of parallelization in the synthesis, there is a trade-off between latency and resource usage for a given algorithm; in these studies, the 25 ns latency requirement is held as a strict upper limit.

These strict constraints motivate several technical choices in the readout algorithm design.
First, only the encoder stage of the VAE is considered for front-end implementation such that the full reconstruction error cannot be calculated on-chip. 
As a result, anomaly detection on-chip must rely on latent space information alone, so the KLD is used as an anomaly discriminant score.
Further, the KLD score is ``clipped" to only include the sum of encoded means in the latent space. 
This serves as an effective substitute for the full KLD at a fraction of the computation cost~\cite{Govorkova_2022}.
This also removes the need for a sampling step, meaning that the track embedded into latent space means is the final compressed track representation that is proposed to be transmitted off-chip.

Additionally, quantization of model weights, biases, and values passing between layers, along with pruning of unnecessary connections, are performed to further reduce the model's computational requirements. 
While the simulated dataset provides input charge values with \texttt{float32} type, the precision of these values on realistic ASICs would likely be significantly limited.
Two cases are considered: a low-precision scenario with input charge value quantization of 4 bits, and a high-precision scenario with 10 bits. 
In the low-precision case, the input charges are mapped into 16 evenly spaced bins ranging from 0 to 8000 with overflow. 
In the high-precision scenario, inputs are mapped to a roughly even distribution between $0$ and $1$ with a preprocessing consisting of $\sqrt{\alpha \cdot \log (x)}$, where $\alpha = 0.094$ is a normalization factor determined by the dataset.


\section{Data Reduction}
\label{sec:data_reduction}

With VAE-based compression, the off-detector data volume is reduced by compressing every event to its smaller latent representation. 
Pixel sensor data is sent off the detector in this lower-dimensional form, and then decompressed by the VAE decoder off-chip, where it can be used by later stages of the DAQ pipeline for further analysis. 
Since the VAE is an inherently unsupervised ML model, it can be trained directly over experimental data, mitigating any reliance on simulation. 
This data reduction scheme can be compared to a classification approach demonstrated with the smart pixel dataset~\cite{yoo2023_on_sensor_filtering}, where the off-detector data rate is reduced via the filtering of low-momentum tracks. 

Each track is modeled by 105 features: 13 pixel charges for 8 time slices and $y_0$, the distance of the charge from the interaction point in the $y$ direction. 
However, the VAE here only uses 53 of those input features, corresponding to the 13 pixel charges of the first 4 time slices and $y_0$.
This input reduction was found to greatly reduce model size with minimal impact on track reconstruction performance from the latent space. 
To retain a full description of the data, such a VAE must encode this limited set of 53 input features to the latent space, and then recreate the original 105 input features, extrapolating to the remaining 4 time slices of data. 

The VAE model has an input dimension of 105, and output dimension of 105, and is constructed with a encoder with hidden layer of size 24, latent space dimension of 8, and decoder consisting of three hidden layers with sizes 24, 48, and 64. 
Layer sizes and design were subject to an optimization procedure designed to optimize performance while maintaining the latency and resource requirements established in Section~\ref{sec:methods}. 
The weights, biases, and activation functions are quantization-aware and given 8 bits of precision.
The model is trained with the \texttt{Adam} optimizer with learning rate 0.001, batch size 1024, and a scale factor for the KLD term $\beta$ which is annealed sinusoidally between $0.00005$ and $0.00025$ during training.
The model is trained for 150 epochs of 440,000 simulated tracks from the smart pixel dataset. 
Then, the encoder is slowly pruned to 0.4 sparsity over 10 epochs, and trained for 50 additional epochs. 

After training, the resulting VAE was evaluated over a test set of 55,000 tracks, resulting in a set of VAE-generated tracks that would constitute the final data in the case of VAE-based compression at the front-end. 
After encoding and off-detector transmission, the decompressed data must be a sufficiently high-fidelity representation of the original track to maintain the same event analysis performance down the line in the DAQ process.
Events that are poorly reconstructed by the VAE can be flagged as anomalous on-chip, and sent off-detector uncompressed for further analysis, as described in Section \ref{sec:anomaly_detection}.

Finally, the VAE is synthesized to estimate its performance, resource usage, and latency in hardware. 
This synthesis is performed with the \textsc{hls4ml} library with a reuse factor of 1 and the \texttt{io\_parallel} strategy for lowest latency.
As discussed in Section \ref{sec:methods}, only the encoder is synthesized wih an assumed output of the track encoded into the latent space mean values. 
The resources considered are those of an FPGA, particularly the digital signal processors (DSPs) commonly used for multiply-and-accumulate operations, look-up tables (LUTs) commonly used for small bit-width calculations, and flip-flops (FFs) for single bit storage.
In both low- and high-precision scenarios, the on-chip encoder is found to evaluate in less than 25 ns using approximately 30,000 LUTs and fewer than 700 DSPs. 

The performance of the VAE for the data compression task is assessed through two criteria: its ability to provide a high quality reconstruction of the input, and the performance of a simple pileup classification task using the VAE-reconstructed data to represent a typical physics analysis task. 
Details of each check are provided below. 

\subsection{VAE Reconstruction Quality}
\begin{figure}[!htbp]
\centering
\includegraphics[width=0.7\textwidth]{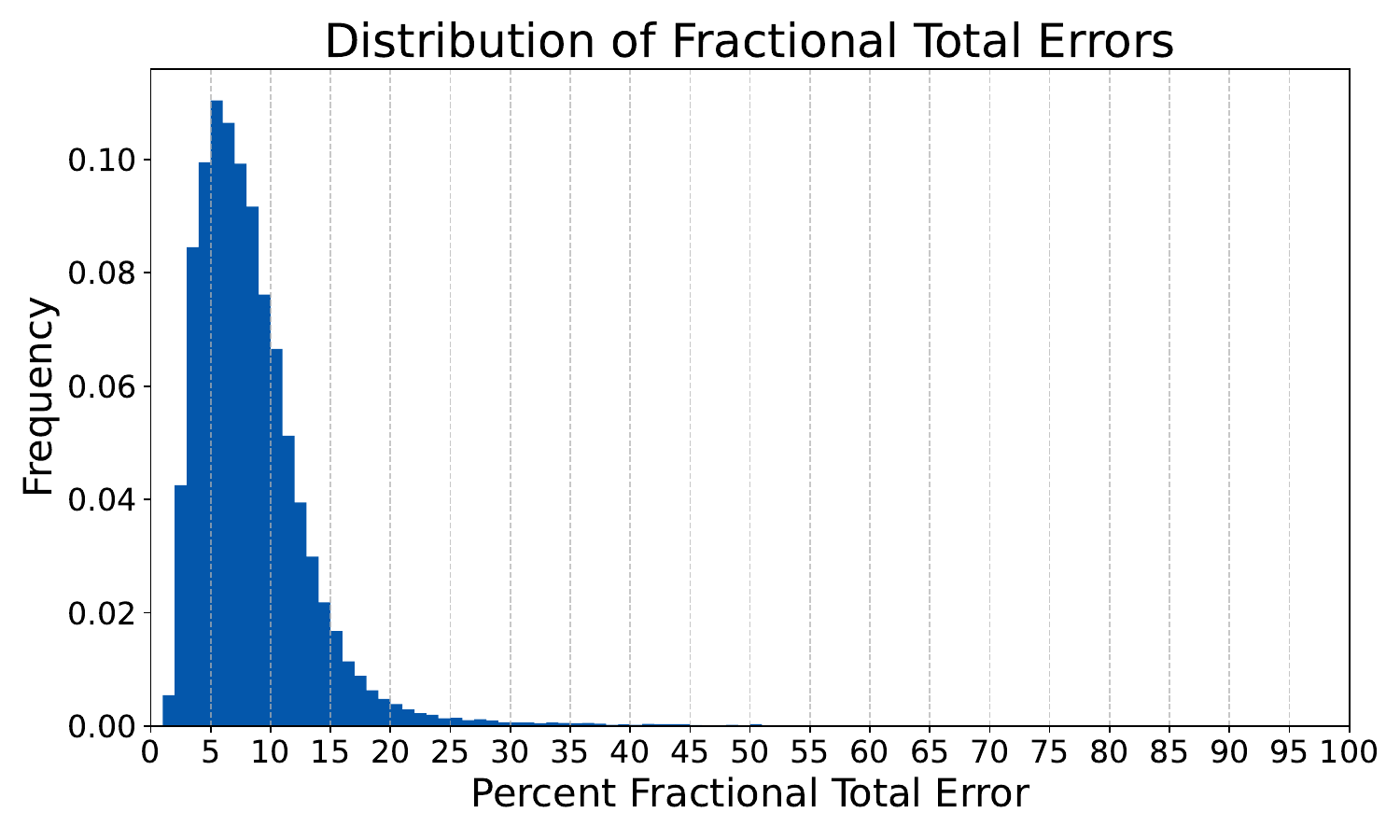}
\caption{ Distribution of the total percent error between the VAE recreated output and truth input, calculated for a given track by summing pixel charge values across the entire sensor, for the low-precision scenario. For a 4-bit quantized input, the VAE is capable of generating output tracks with a mean reconstruction error of 8.5\% and median reconstruction error of 7.5\%. 
\label{fig:total_percent_error}}
\end{figure}

To quantify the difference between the reconstructed and truth input quantities, we define a metric, the ``fractional total error,'' as follows: $Er = \frac{\sum_i{|q_{r_i} - q_i|}}{\sum_i{q_i}}$, where $q_{ri}$ is the reconstructed charge for pixel $i$ and $q_i$ is the actual charge for pixel $i$.
The distribution of this metric across the test set of tracks is presented in Figure~\ref{fig:total_percent_error}.

\begin{figure}[!htbp]
\centering
\includegraphics[width=0.95\textwidth]{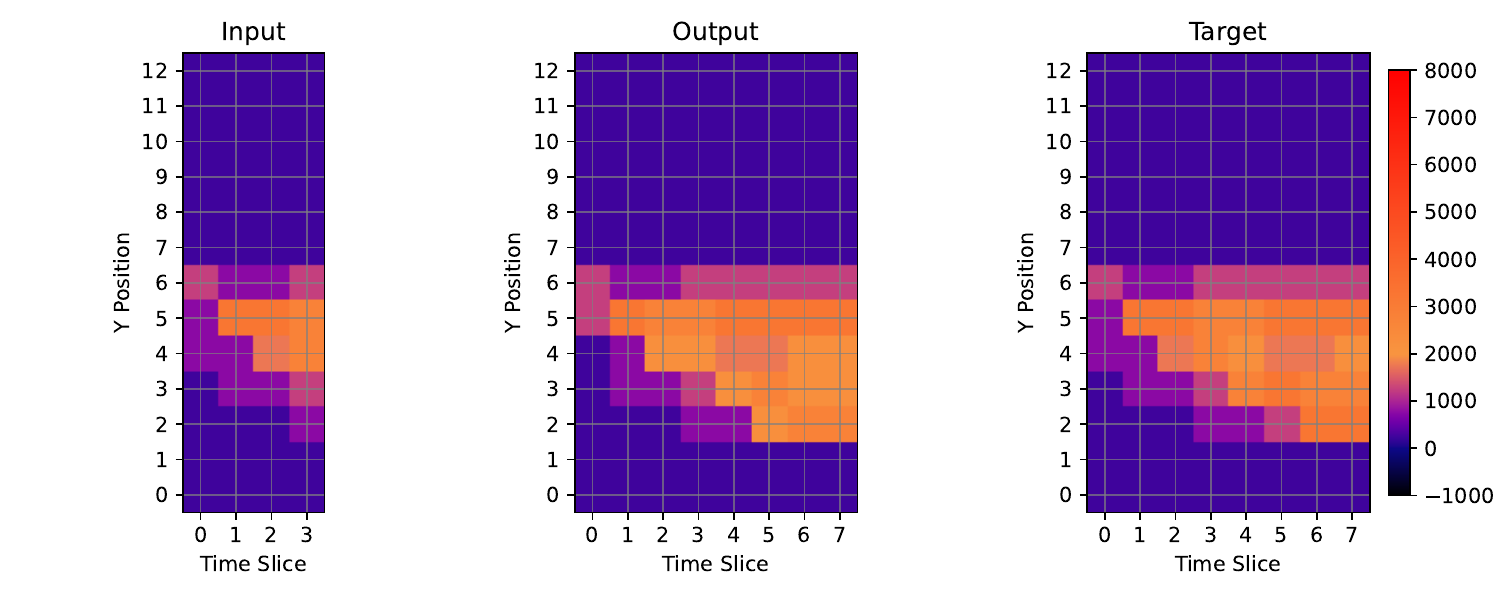}
\caption{Example high $p_\text{T}$ track from the smart pixel dataset, displaying the truth VAE input of 4 time slices, (left), VAE reconstructed output (middle), and target truth charge distribution across all 8 time slices (right). The summed sensor error for this example is 9.3\%. 
The VAE demonstrates a knowledge of the underlying charged particle process that is sufficient to extrapolate the four input time slices to a trajectory that is consistent with that of the truth deposited charge. 
\label{fig:example_reco}}
\end{figure}

The mean percent error on the summed sensor charge is 15.3\% in the high-precision scenario (10 bit pixel values) and 8.5\% in the low-precision scenario (4 bit pixel values). 
Reconstructed tracks generally preserve the position and distribution integrity of the original track. 
Figure~\ref{fig:example_reco} provides an example high $p_\text{T}$ track with a sum reconstruction error of 9.3\%, showing the input (truncated to four time slices), along with the VAE-reconstructed output and truth charge values for all eight time slices.
The ability of the VAE to extrapolate the additional four time slices and generate a faithful representation of the original data is presumed to arise from its intelligent embedding of sensor-wide information to the 8 latent dimensions, a sufficient size to well-describe the underlying process of charged particles in a magnetic field. 

\subsection{VAE-Reconstructed Data for Pileup Classification}

To further assess the quality of the VAE-reconstructed data after compression, we define a simple classification task using the truth labels in the smart pixel dataset: distinguish between high-$p_\text{T}$ tracks ($p_\text{T} \ge$ 2 GeV) and lower $p_\text{T}$ tracks ($p_\text{T} <$ 2 GeV) based on the pattern of deposited charge in the pixel sensor. 
This is a balanced classification task as the high $p_\text{T}$ tracks make up 50\% of the dataset.
Such classification provides physically useful information as high $p_\text{T}$ tracks likely arise from the primary vertex, whereas lower $p_\text{T}$ are likely pileup tracks.
Specifically, the uncompressed representation of each track as generated by the VAE's decoder stage is given as input to an off-detector classifier optimized for this classification task. 
The on-chip classifier was subject to an optimization procedure to ensure the highest achievable performance within the establish latency and resource constraints of the on-chip environment. 
Two signal efficiency (true positive rate) working points are studied, 93\% and 98\%, though it is anticipated that in a realistic detector design, the signal efficiency will need to be very high to minimize any loss of valuable collision data. 
Any imperfection in reconstructed data could be evaluated through dedicated calibration studies and incorporated into final physics measurements as a systematic uncertainty.

This scheme is capable of achieving a background rejection (true negative rate) of 10\% (21\%) for a fixed signal efficiency of 98\% (93\%) in the low-precision scenario.
Such performance is in line with previous on-chip classification studies~\cite{yoo2023_on_sensor_filtering}, indicating good capability for offline physics analysis using the VAE-reconstructed sensor information. 
As the 104 4-bit pixel sensor inputs are proposed to be sent off-detector in the latent space representation of eight dimensions with 8-bit precision, the expected compressed size is 15.4\% of an uncompressed representation for data transmitted off the detector.
This represents a decrease in off-detector data rate that is achievable while maintaining good physics performance of reconstructed data.


To compare these results to other front-end ML compression solutions, an alternate model is designed to simply perform the high vs. low $p_\text{T}$ track classification directly on the chip, in line with the approach described in Ref.~\cite{yoo2023_on_sensor_filtering}.
The on-chip classifier model takes as inputs the charge depositions in the 13 rows of pixels in the $y$ axis for each of the first 4 time slices, concatenated with $y_0$ for 53 total input features. 
The tracks modeled by these inputs are given to a deep neural network with two hidden layers of sizes 24 and 12 respectively. 
Each layer is followed by a ReLU activation function and has weights and biases quantized to 10 bits. 
The output of the model is a single predictive score of whether the event is a high $p_\text{T}$ or low $p_\text{T}$ event. 
The model is trained with the \texttt{Adam} optimizer with learning rate 0.001 and batch size 1024 over 100 epochs of the 440,000 simulated tracks from the smart pixel dataset. 
After the initial training, the model is pruned down to 0.4 sparsity over 10 epochs and then trained for 50 additional epochs.

Table~\ref{tab:model_metrics} provides a comparison of key model metrics for both the VAE and the on-chip classifier discussed above, namely the latency; number of LUTs, DSPs, and FFs used; background rejection for two fixed signal efficiency working points as a proxy for physics performance; and off-detector data rate reduction. 
The data compression for the classifier is calculated by assuming tracks classified as background will not be transmitted and removing tracks from the off-detector data in accordance with the background rejection rate. 

The VAE scheme, with an on-chip encoder followed by an off-detector decoder and classifier, meets or outperforms the on-chip classifier approach in all metrics, with a lower latency, lower or similar resource usage, and lower off-detector data rate, along with better performance of the pileup classification task. 
This indicates the benefit of the VAE-based compression approach as compared to a direct application-specific model for filtering, motivating the relocation of this task to the edge. 
Furthermore, this algorithm for data reduction is agnostic to the type of information received from the detector, and thus could be applied to a wide range of detector subsystems. 
Coupled with the order of magnitude reduction in off-detector data rate from the VAE compression method, this indicates the method's broad potential to enhance future DAQ schemes.

\begin{table}[!htbp]
\centering   

\begin{tabular}{|c|c|c|c|c|}
    \hline
    & \multicolumn{2}{c|}{\textbf{High-Precision Scenario}} & \multicolumn{2}{c|}{\textbf{Low-Precision Scenario}} \\
    \hline
    \makecell{}& \makecell{VAE (+ Off-Chip \\ Classifier)} & \makecell{On-Chip \\ Classifier} & \makecell{VAE (+ Off-Chip \\ Classifier)} & \makecell{On-Chip \\ Classifier} \\
    \hline
    Latency [ns] & \textbf{15} & 25 & \textbf{15} & 20 \\
    \hline
    LUTs & 28,653 & 38,394 & 33,252 & 32,903 \\
    \hline   
    DSPs & \textbf{652} & 723 & \textbf{378} & 798 \\
    \hline   
    FFs & 850 & 931 & 1043 & 906 \\
    \hline   
    BR For SE = 93\% & \textbf{38\%} & 32\% & \textbf{21\%} & 18\% \\
    \hline 
    BR For SE = 98\% & \textbf{25\%} & 18\% & \textbf{10\%} & 7\% \\
    \hline
    Compressed Size & \textbf{7.6\%} & 82\% & \textbf{15.4\%} & 93\% \\
    \hline
\end{tabular}
\caption{
    Summary of model performance metrics, namely latency, on-detector resources (LUTs, DSPs, FFs), background rejection (BR) for two fixed signal efficiencies (SE) on the pileup classification task, and the compressed size, given as a percent of the original data volume that is transmitted off the detector. Two models are shown: the VAE scheme which includes an on-detector encoder followed by off-detector decoder and classifier stages, and a classifier that can fit on-detector requirements.
    Resources quoted only refer to the architecture components that are proposed to be deployed on-chip. 
}
\label{tab:model_metrics}
\end{table}


\section{Anomaly Detection} 
\label{sec:anomaly_detection}

While VAE-based data compression is a traditional application, this result extends to the novel capability of the encoder to detect anomalies at the edge using the track's latent space encoding. 
The data-driven training of the VAE allows the model to learn the underlying data distribution and detect outlier events without the use of a signal model, enabling agnostic detection of unusual detector activity ranging from detector malfunctions~\cite{dq2024} to potential new physics. 
Furthermore, the ability to detect anomalies in real-time through a VAE-based readout scheme could allow the readout to dynamically decide track-by-track whether to compress the output or transmit it in full uncompressed precision for further study. 
Since outlier events are often the most challenging for the VAE to accurately reconstruct, transmitting these events at full precision off-detector could help mitigate the loss of key information of anomalous pixel events. 

\begin{figure}[!htbp]
\centering
\includegraphics[width=0.95\textwidth]{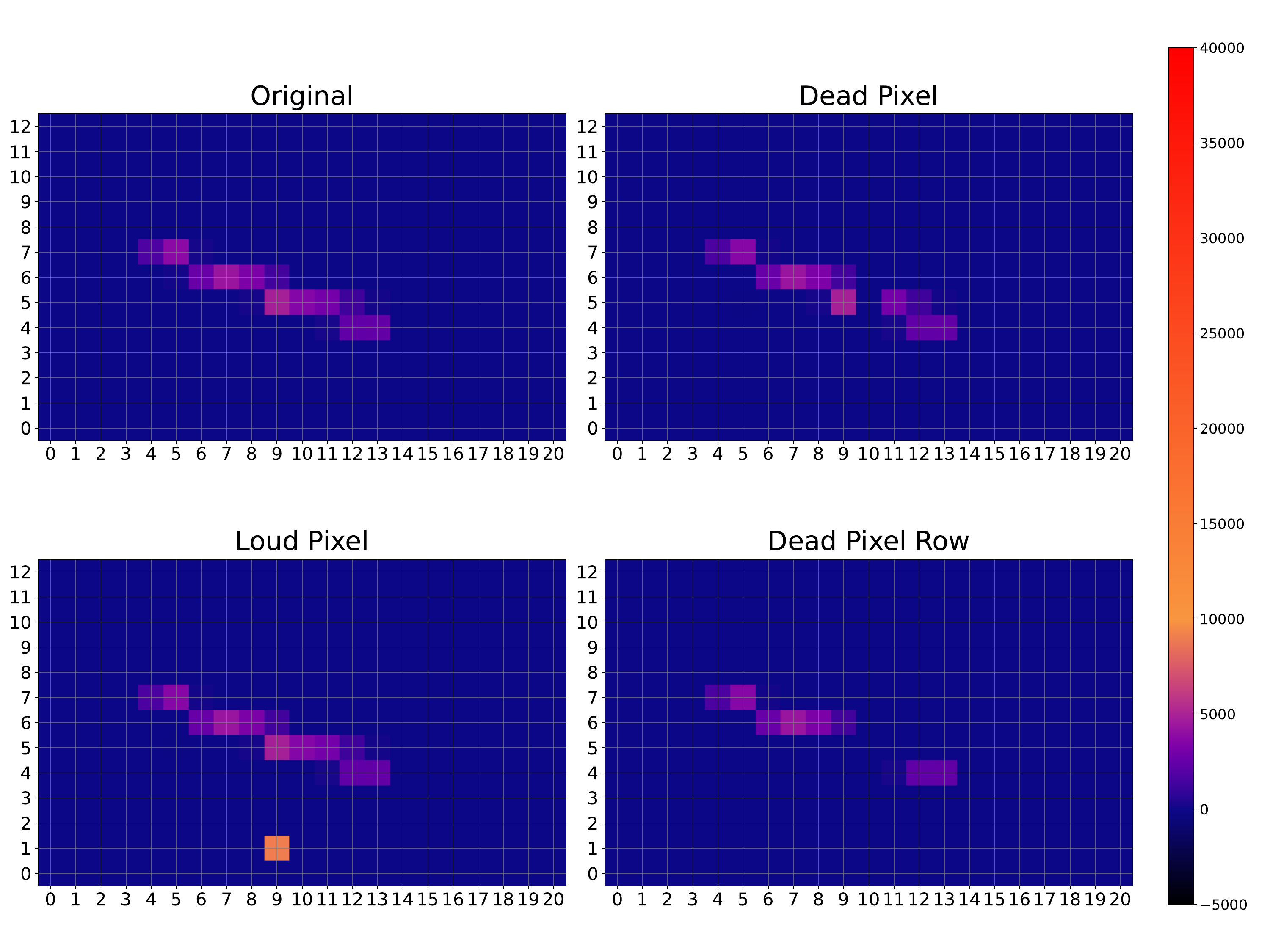}
\caption{
    Display of the typical simulated tracks and their pattern of charge deposition across the sensor from the smart pixel dataset, including a typical background track (top left), along with the three types of simulated anomalies, namely a dead pixel (top right), loud pixel (bottom left), and a dead pixel row (bottom right). 
}
\label{fig:example_tracks}
\end{figure}

To test the anomaly detection capability of the VAE described in Section~\ref{sec:data_reduction}, synthetic pixel events are constructed to represent various detector anomalies, starting from the original smart pixel tracks in the test set.
Three different kinds of detector anomalies are considered: sensors with a single dead pixel, sensors with a single noisy pixel, or sensors with a dead row of pixels. 
These anomalies represent possible detector malfunctions that are important to identify and flag.  
Individual dead pixels are randomly chosen from the brighter half of pixels with nonzero charge.
Dead pixels rows are randomly chosen from pixel rows with nonzero charge and their charge is set to a value of zero. 
The charge of a loud pixel in the loud pixel anomaly is 9000, determined by the 99th percentile of the max pixel value per track across the entire dataset. 
Figure~\ref{fig:example_tracks} displays these anomalies as they appear in the pixel sensor, alongside a standard smart pixel track for reference. 
These synthetic anomalous tracks undergo the same input processing as the standard tracks in both the low- and high-precision scenarios.

\begin{figure}[!htbp]
\centering
   \includegraphics[width=0.7\textwidth]{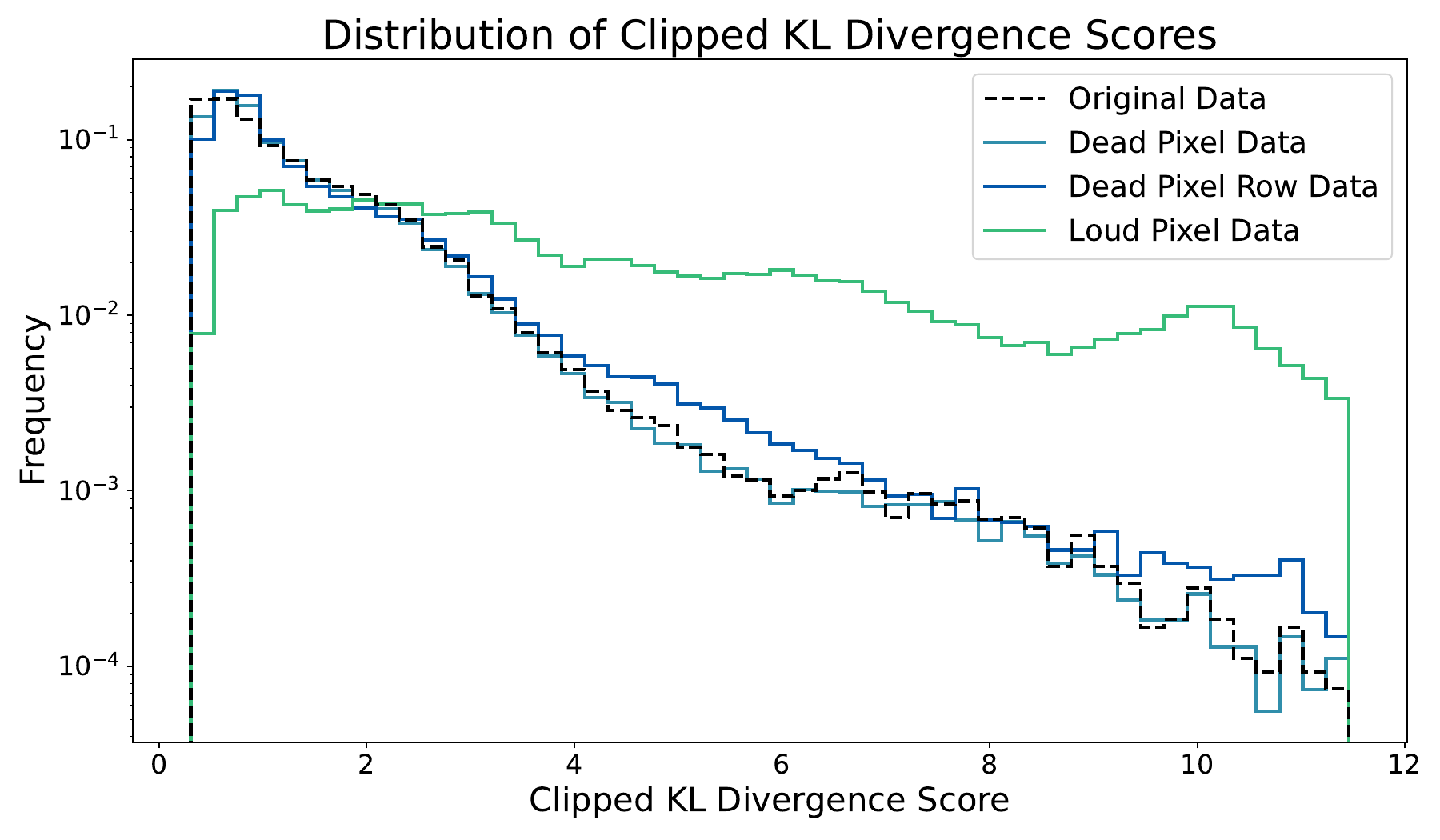}
    \caption{Log scale distribution of the anomaly score, defined as the ``clipped" KLD, for the original data compared to the three classes of synthetic anomalies. Loud pixel and dead pixel row anomalies are found to have a higher mean anomaly score than the non-anomalous data, indicating the utility of the VAE's latent space information to perform real-time anomaly detection. 
    \label{fig:kld_distributions}}
\end{figure}

The encoder of the same VAE described in the previous section is evaluated over test sets of approximately 55,000 events for each of the three simulated anomalies, and the clipped KLD scores are calculated for each event for use as an anomaly classifier score. 
Figure~\ref{fig:kld_distributions} displays the anomaly score distributions of the standard track and anomaly test sets. 
Each of the anomalies is reconstructed with a higher mean anomaly score compared to the standard tracks, with a tail to higher values indicating a higher likelihood of incompatibility with the background. 
The anomaly detection capability was found to be anticorrelated with the reconstruction quality, indicating an area in need of further study and optimization for a realistic front-end scheme. 
Table~\ref{tab:anomaly_detection_metrics} shows the achievable efficiency for identifying sensor anomalies for a fixed background track rejection of 90\%. 

ML at any stage of collider DAQ systems must be implemented carefully to avoid alteration or corruption of the raw data. 
While VAE-based sensor data compression can mitigate high data rates during ordinary operation with minimal artifacts introduced in reconstruction, any indication of anomalous behavior in the sensors would ideally be preserved in original form. 
Thus, the compression and anomaly detection capabilities shown here could be used in tandem in a front-end readout design so that the events with the highest anomaly score are flagged and sent off the detector with full precision. 
With 1\% of tracks with the highest anomaly score being transmitted uncompressed, this scheme could still achieve a data rate reduction of 16.4\% of the original size while preserving the most unusual detector behaviors at full precision for offline study. 

\begin{table}[!htbp]
\centering   
\begin{tabular}{|c|c|c|c|}
    \hline
    \multicolumn{4}{|c|}{Signal Efficiency for 90\% Background Rejection} \\
    \hline
     & Dead Pixel & Dead Pixel Row & Loud Pixel  \\
    \hline
    VAE (Low Precision) & 10\% & 12.9\% & 58.5\% \\
    \hline
    VAE (High Precision) & 10.8\% & 17.0\% & 22.7\% \\
    \hline
\end{tabular}
\caption{
    Signal efficiency for the three synthetic pixel anomalies for a fixed background rejection of 90\%, for both the low- and high-precision scenarios.
}
\label{tab:anomaly_detection_metrics}
\end{table}


\section{Conclusions}
\label{sec:conclusions}

A VAE capable of performing both data compression and anomaly detection in on-detector silicon pixel readout is presented. 
With an on-chip encoder and off-detector decompression, a VAE-based readout approach can reduce the off-detector data rate to 15.4\% of its original size for 4-bit charge values, while providing a faithful reconstruction of the original pixel charge distribution.
Further, information from the VAE latent space can enable anomaly detection with potential for real-time monitoring of pixel data quality. 
Both of these capabilities can be achieved with a latency less than 25 ns, better performance, and similar or fewer resources than a comparable on-chip track filtering approach.
This results opens the door to intelligent inference in front-end readout of modern collider experiments, with potential to dramatically improve the real-time filtering systems required to manage collider data rates, and via a generic architecture that can be applied to a wide variety of detector subsystems.  

Future work will seek to further test the fidelity of the decompressed data, specifically working to further reduce the reconstruction error and to study more complex reconstruction tasks such as feature regression or pattern recognition.
The resource constraints in a real detector deployment will be strongly dependent on the power requirements of this system, requiring careful evaluation of on-chip power needs and tailoring to the specific application. 
Additional comparison to traditional on-chip data compression approaches~\cite{Garcia-Sciveres_2014, Garcia-Sciveres_2014_2, GARCIASCIVERES201618} will be essential motivation for the broad introduction of ML-based readout methodologies. 
The generality of this approach can also be leveraged to consider the integration of front-end VAEs into a variety of other detector technologies, such as for waveform analysis of calorimeter or drift chamber signals.
A key step will be the assessment of potential hardware targets for these models and ensuring their viability for deployment in the extreme collider environment.


\section*{Acknowledgements}

This work is supported by the U.S. Department of Energy under contract number DE-AC02-76SF00515 and the Office of the Vice Provost for Undergraduate Education at Stanford University.




\section*{References}

\bibliographystyle{unsrt}
\bibliography{main}

\begin{thebibliography}{10}

\bibitem{p5_2023}
{H. Murayama, K. Heeger, et al}.
\newblock {Exploring the Quantum Universe: Pathways to Innovation and Discovery
  in Particle Physics}, Dec 2023.
\newblock https://www.usparticlephysics.org/2023-p5-report/.

\bibitem{doe_brn}
{Basic Research Needs for High Energy Physics Detector Research \&
  Development}, Dec 2019.
\newblock https://science.osti.gov/-/media/hep/pdf/Reports/2020/ \\
  DOE\_Basic\_Research\_Needs\_Study\_on\_High\_Energy\_Physics.pdf.

\bibitem{apresyan2023detector}
A.~Apresyan, M.~Artuso, J.~Brau, H.~Chen, M.~Demarteau, Z.~Demiragli, S.~Eno,
  J.~Gonski, P.~Grannis, H.~Gray, O.~Gutsche, C.~Haber, M.~Hohlmann,
  J.~Hirschauer, G.~Iakovidis, K.~Jakobs, A.~J. Lankford, C.~Pena,
  S.~Rajagopalan, J.~Strube, C.~Tully, C.~Vernieri, A.~White, G.~W. Wilson,
  S.~Xie, Z.~Ye, J.~Zhang, and B.~Zhou.
\newblock {Detector R\&D needs for the next generation $e^+e^-$ collider},
  2023.

\bibitem{duarte2022fastmlsciencebenchmarksaccelerating}
Javier Duarte, Nhan Tran, Ben Hawks, Christian Herwig, Jules Muhizi, Shvetank
  Prakash, and Vijay~Janapa Reddi.
\newblock {FastML Science Benchmarks: Accelerating Real-Time Scientific Edge
  Machine Learning}, 2022.
\newblock https://arxiv.org/abs/2207.07958.

\bibitem{carini2022smartsensor}
Gabriella Carini, Grzegorz Deptuch, Jennet Dickinson, Dionisio Doering, Angelo
  Dragone, Farah Fahim, Philip Harris, Ryan Herbst, Christian Herwig, Jin
  Huang, Soumyajit Mandal, Cristina~Mantilla Suarez, Allison~McCarn Deiana,
  Sandeep Miryala, F.~Mitchell Newcomer, Benjamin Parpillon, Veljko Radeka,
  Dylan Rankin, Yihui Ren, Lorenzo Rota, Larry Ruckman, and Nhan Tran.
\newblock {Smart sensors using artificial intelligence for on-detector
  electronics and ASICs}, 2022.
\newblock {FERMILAB-CONF-22-430-PPD-SCD}.

\bibitem{kulkarni2023onsensordatafilteringusing}
Shruti~R. Kulkarni, Aaron Young, Prasanna Date, Narasinga~Rao Miniskar,
  Jeffrey~S. Vetter, Farah Fahim, Benjamin Parpillon, Jennet Dickinson, Nhan
  Tran, Jieun Yoo, Corrinne Mills, Morris Swartz, Petar Maksimovic,
  Catherine~D. Schuman, and Alice Bean.
\newblock {On-Sensor Data Filtering using Neuromorphic Computing for High
  Energy Physics Experiments}.
\newblock In {\em Proceedings of On-Sensor Data Filtering using Neuromorphic
  Computing for High Energy Physics (ICONS '23)}, 2023.
\newblock https://arxiv.org/abs/2307.11242.

\bibitem{Gonski_2024}
J.~Gonski, A.~Gupta, H.~Jia, H.~Kim, L.~Rota, L.~Ruckman, A.~Dragone, and
  R.~Herbst.
\newblock {Embedded FPGA developments in 130 nm and 28 nm CMOS for machine
  learning in particle detector readout}.
\newblock {\em Journal of Instrumentation}, 19(08):P08023, August 2024.

\bibitem{Johnson2024_ngc_eFPGAs}
Jyothisraj Johnson, Billy Boxer, Tarun Prakash, Carl Grace, Peter Sorensen, and
  Mani Tripathi.
\newblock {Investigating resource-efficient neutron/gamma classification ML
  models targeting eFPGAs}.
\newblock {\em Journal of Instrumentation}, 19(07):P07034, July 2024.

\bibitem{10.3389/fdata.2022.787421}
Allison~McCarn Deiana, Nhan Tran, Joshua Agar, Michaela Blott, Giuseppe
  Di~Guglielmo, Javier Duarte, Philip Harris, Scott Hauck, Mia Liu, Mark~S.
  Neubauer, Jennifer Ngadiuba, Seda Ogrenci-Memik, Maurizio Pierini, Thea
  Aarrestad, Steffen Bähr, Jürgen Becker, Anne-Sophie Berthold, Richard~J.
  Bonventre, Tomás~E. Müller~Bravo, Markus Diefenthaler, Zhen Dong, Nick
  Fritzsche, Amir Gholami, Ekaterina Govorkova, Dongning Guo, Kyle~J.
  Hazelwood, Christian Herwig, Babar Khan, Sehoon Kim, Thomas Klijnsma, Yaling
  Liu, Kin~Ho Lo, Tri Nguyen, Gianantonio Pezzullo, Seyedramin Rasoulinezhad,
  Ryan~A. Rivera, Kate Scholberg, Justin Selig, Sougata Sen, Dmitri Strukov,
  William Tang, Savannah Thais, Kai~Lukas Unger, Ricardo Vilalta, Belina von
  Krosigk, Shen Wang, and Thomas~K. Warburton.
\newblock {Applications and Techniques for Fast Machine Learning in Science}.
\newblock {\em Frontiers in Big Data}, 5, 2022.

\bibitem{yoo2023_on_sensor_filtering}
Jieun Yoo, Jennet Dickinson, Morris Swartz, Giuseppe~Di Guglielmo, Alice Bean,
  Douglas Berry, Manuel~Blanco Valentin, Karri DiPetrillo, Farah Fahim, Lindsey
  Gray, James Hirschauer, Shruti~R. Kulkarni, Ron Lipton, Petar Maksimovic,
  Corrinne Mills, Mark~S. Neubauer, Benjamin Parpillon, Gauri Pradhan, Chinar
  Syal, Nhan Tran, Dahai Wen, and Aaron Young.
\newblock {Smart pixel sensors: towards on-sensor filtering of pixel clusters
  with deep learning}.
\newblock {\em Mach. Learn.: Sci. Technol.}, 5:035047, 2024.

\bibitem{dickinson2023_on_sensor_inference}
Jennet Dickinson, Rachel Kovach-Fuentes, Lindsey Gray, Morris Swartz,
  Giuseppe~Di Guglielmo, Alice Bean, Doug Berry, Manuel~Blanco Valentin, Karri
  DiPetrillo, Farah Fahim, James Hirschauer, Shruti~R. Kulkarni, Ron Lipton,
  Petar Maksimovic, Corrinne Mills, Mark~S. Neubauer, Benjamin Parpillon, Gauri
  Pradhan, Chinar Syal, Nhan Tran, Dahai Wen, Jieun Yoo, and Aaron Young.
\newblock Smartpixels: Towards on-sensor inference of charged particle track
  parameters and uncertainties, 2023.
\newblock https://arxiv.org/abs/2312.11676.

\bibitem{Dickinson2022_Smart_Pixel_Dataset}
Morris Swartz and Jennet Dickinson.
\newblock Smart pixel dataset, 2022.
\newblock https://zenodo.org/records/7331128.

\bibitem{9447722}
Giuseppe~Di Guglielmo, Farah Fahim, Christian Herwig, Manuel~Blanco Valentin,
  Javier Duarte, Cristian Gingu, Philip Harris, James Hirschauer, Martin Kwok,
  Vladimir Loncar, Yingyi Luo, Llovizna Miranda, Jennifer Ngadiuba, Daniel
  Noonan, Seda Ogrenci-Memik, Maurizio Pierini, Sioni Summers, and Nhan Tran.
\newblock {A Reconfigurable Neural Network ASIC for Detector Front-End Data
  Compression at the HL-LHC}.
\newblock {\em IEEE Transactions on Nuclear Science}, 68(8):2179--2186, 2021.

\bibitem{Shenoy_2023}
Rohan Shenoy, Javier Duarte, Christian Herwig, James Hirschauer, Daniel Noonan,
  Maurizio Pierini, Nhan Tran, and Cristina Mantilla~Suarez.
\newblock {Differentiable Earth mover’s distance for data compression at the
  high-luminosity LHC}.
\newblock {\em Machine Learning: Science and Technology}, 4(4):045058, December
  2023.

\bibitem{kingma2022autoencodingvariationalbayes}
Diederik~P Kingma and Max Welling.
\newblock {Auto-Encoding Variational Bayes}.
\newblock 2022.
\newblock https://arxiv.org/abs/1312.6114.

\bibitem{BELIS2024100091}
Vasilis Belis, Patrick Odagiu, and Thea~Klaeboe Aarrestad.
\newblock Machine learning for anomaly detection in particle physics.
\newblock {\em Reviews in Physics}, 12:100091, 2024.

\bibitem{CMS:2008xjf}
S.~Chatrchyan et~al.
\newblock {The CMS Experiment at the CERN LHC}.
\newblock {\em JINST}, 3:S08004, 2008.

\bibitem{Duarte:2018ite}
Javier Duarte et~al.
\newblock {Fast inference of deep neural networks in FPGAs for particle
  physics}.
\newblock {\em JINST}, 13(07):P07027, 2018.

\bibitem{fastml_hls4ml}
{FastML Team}.
\newblock fastmachinelearning/hls4ml, 2023.
\newblock https://github.com/fastmachinelearning/hls4ml.

\bibitem{Govorkova_2022}
Ekaterina Govorkova, Ema Puljak, Thea Aarrestad, Thomas James, Vladimir Loncar,
  Maurizio Pierini, Adrian~Alan Pol, Nicolò Ghielmetti, Maksymilian Graczyk,
  Sioni Summers, Jennifer Ngadiuba, Thong~Q. Nguyen, Javier Duarte, and Zhenbin
  Wu.
\newblock Autoencoders on field-programmable gate arrays for real-time,
  unsupervised new physics detection at 40 mhz at the large hadron collider.
\newblock {\em Nature Machine Intelligence}, 4(2):154–161, February 2022.

\bibitem{dq2024}
{Abadjiev, D. et al}.
\newblock {Autoencoder-Based Anomaly Detection System for Online Data Quality
  Monitoring of the CMS Electromagnetic Calorimeter}.
\newblock {\em Computing and Software for Big Science}, 8(1), June 2024.

\bibitem{Garcia-Sciveres_2014}
M~Garcia-Sciveres and X~Wang.
\newblock Data compression considerations for detectors with local
  intelligence.
\newblock {\em Journal of Instrumentation}, 9(10):C10011, oct 2014.

\bibitem{Garcia-Sciveres_2014_2}
M~Garcia-Sciveres and X~Wang.
\newblock Data encoding efficiency in binary strip detector readout.
\newblock {\em Journal of Instrumentation}, 9(04):P04021, apr 2014.

\bibitem{GARCIASCIVERES201618}
Maurice Garcia-Sciveres and Xinkang Wang.
\newblock Data encoding efficiency in pixel detector readout with charge
  information.
\newblock {\em Nuclear Instruments and Methods in Physics Research Section A:
  Accelerators, Spectrometers, Detectors and Associated Equipment}, 815:18--22,
  2016.

\end{thebibliography}

\end{document}